\documentclass[]{spie}  %>>> use for US letter paper
\usepackage{amsmath,amsfonts,amssymb}
\usepackage{graphicx}
\usepackage[colorlinks=true, allcolors=blue]{hyperref}
\usepackage{subfig}

\title{Closed-loop Atmospheric Dispersion Correction for High-Contrast Imaging with MagAO-X\footnote{\hspace{.3 cm}This paper includes data gathered with the 6.5 meter Magellan Telescopes located at Las Campanas Observatory, Chile.}}

\author[a]{Katie Twitchell}
\author[b,c]{Sebastiaan Haffert}
\author[b]{Jared R. Males}
\author[b]{Laird M. Close}
\author[a,b,d]{Olivier Guyon}
\author[b]{Kyle Van Gorkom}
\author[e]{Alexander Hedglen}
\author[b]{Parker T. Johnson}
\author[a]{Maggie Y. Kautz}
\author[a]{Jay K. Kueny}
\author[a]{Joshua Liberman}
\author[b,d]{Miles Lucas}
\author[h]{Avalon McLeod}
\author[c]{Matthijs Mars}
\author[a]{Eden A. McEwen}
\author[b]{Jialin Li}
\author[g]{Joseph D. Long}
\author[a]{Jhen Lumbres}
\author[f]{Lauren Schatz}
\author[c]{Elena Tonucci}

\affil[a]{Wyant College of Optical Sciences, University of Arizona, 1630 E University Boulevard, Tucson, Arizona}
\affil[b]{Steward Observatory, University of Arizona, 933 N Cherry Ave, Tucson, Arizona}
\affil[c]{Leiden Observatory, Leiden University, Einsteinweg 55, Leiden, The Netherlands}
\affil[d]{Subaru Telescope, National Observatory of Japan, National Institutes of Natural Sciences, 650 N. A'ohoku Place, Hilo, Hawai'i}
\affil[e]{Northrop Grumman Corporation, 600 South Hicks Road, Rolling Meadows, Illinois}
\affil[f]{Starfire Optical Range, Kirtland Air Force Base, Albuquerque, New Mexico}
\affil[g]{Center for Computational Astrophysics, Flatiron Institute, 162 5th Avenue, New York, New York}
\affil[h]{Draper Laboratory, 555 Technology Square, Cambridge, Massachusetts}

\authorinfo{Further author information: (Send correspondence to K.T.)\\K.T.: E-mail: twitchell@arizona.edu}

\begin{document} 
\maketitle

\begin{abstract}
Incoming starlight is refracted as it enters Earth’s atmosphere from the vacuum of space. The wavelength-dependence of atmospheric refraction causes elongation of the broadband PSF of ground-based telescopes, especially in the visible spectrum. The result is degraded image quality alongside reduced coronagraph light-blocking efficiency, both of which limit high-contrast observations. An atmospheric dispersion corrector (ADC) is a dispersive optic used to compensate for this effect. Current methods for dispersion compensation use analytical models to anticipate dispersion strength based on parameters such as site altitude and telescope zenith angle; however, dispersion strength is also dictated by a number of factors that cannot be measured, including instantaneous humidity, temperature, and pressure along the line of sight to the star. This leads to constant over- or under-correction of the true atmospheric dispersion by the ADC. In this work, we use the Magellan extreme adaptive optics system MagAO-X at Las Campanas Observatory to measure and correct residual atmospheric dispersion in real-time. The amount of residual dispersion is encoded in the orientation of satellite spots generated by using MagAO-X’s deformable mirror as a diffraction grating. We have used these real-time measurements as feedback for closed-loop control of the ADCs on-sky at visible and NIR wavelengths, reducing residual atmospheric dispersion down to sub-mas$/\mu$m levels. Active atmospheric dispersion correction on MagAO-X is a precursor to high-contrast imaging with Extreme AO for the upcoming Extremely Large Telescopes, where high-precision dispersion compensation will be required to image exoplanets in reflected light.
\end{abstract}


\section{INTRODUCTION}
\label{sec:intro} 

When a ground-based telescope is pointed at a non-zenith elevation angle, atmospheric refraction occurs. Like most media, the refractive index of air is wavelength-dependent, leading to the phenomenon of atmospheric dispersion\cite{devaney2008chromatic}. In the focal plane, atmospheric dispersion manifests as a wavelength-dependent tip/tilt aberration and causes elongation of the telescope's broadband point-spread function (PSF)\cite{wallner1976effects}. When adaptive optics (AO) is used to enable diffraction-limited imaging from the ground, this elongation becomes a significant source of error and decreases system performance if left uncorrected \cite{correia2020hcipwfs}. The increase in PSF size decreases both the resolution along the dispersion axis and the signal-to-noise ratio (SNR) of objects imaged by smearing photons over a larger number of pixels\cite{owner2004some}. It also affects the behavior of high-precision instruments downstream of the AO system, including coronagraphs or spectrographs, which rely on a high-quality input PSF to achieve optimal performance.\cite{guyon2006coronagraph,wehbe2020impact,wang2020atmospheric}

Wavefront sensing and control in an AO system is targeted towards phase aberrations that affect PSF quality. However, the wavelength-dependent tilt caused by dispersion cannot be corrected by the AO system due to the chromatic nature of dispersion and the achromatic nature of the deformable mirror that performs the wavefront correction. Instead, a dispersive optic must be used. Atmospheric Dispersion Correctors (ADCs) come in many forms, but the most common is a rotational ADC, which comes in the form of a pair of identical prisms placed along the same optical axis.\cite{wynne1986atmospheric} The two prisms, each in their own 360° rotation mount, are controlled with two kinematic degrees of freedom: they either counter-rotate (one clockwise, the other counter-clockwise) to change the strength of dispersion, or they co-rotate as a pair (both clockwise or both counter-clockwise) to change the axis along which the dispersion correction is applied. The prisms are fabricated from glasses with well-known chromatic properties chosen to give the system the proper dynamic range to compensate for dispersion over a range of wavelengths and zenith angles.\cite{pember2020selecting} 

Currently, the standard for ADC control involves an analytical model of the atmosphere.\cite{spano2014accurate} Theoretical atmospheric dispersion as a function of airmass is computed using general observing parameters including site altitude and median temperature, pressure, and humidity. By combining the atmospheric and prism models, a look-up table of ADC rotation angles  that best cancel out atmospheric dispersion for a given telescope zenith angle is generated. However, the accuracy of atmospheric dispersion modeling has been found to be limited\cite{wehbe2021sky}; additionally, the parameters that affect atmospheric refraction are variable and impossible to know precisely along the line of sight of the telescope\cite{devaney2008chromatic}. As a result, the current standard for ADC control leaves some amount of unwanted residual atmospheric dispersion in the PSF. While this residual is inconsequential compared to the diffraction limit of many telescopes, it 
becomes a dominating source of error when considering the large apertures of the new generation of Extremely Large Telescopes (ELTs), motivating the need for closed-loop control.\cite{goncharov2007atmospheric,bahrami2011achromatic}

Previous attempts for higher-precision ADC control have used AO-corrected science camera images to align the ADCs upon their commissioning; however, these methods can only be used as a calibration before the actual observation because they require purposeful displacement of the ADCs, which disrupts the quality of the science image.\cite{cabral2021simple,gao2023measurement} In this work, we propose a method of active ADC control based on a method introduced by the SCEx-AO instrument at the Subaru Telescope that measures residual atmospheric dispersion using diffraction spots in the science focal plane without disrupting ongoing observations.\cite{pathak2016high,pathak2018sky} After the first-stage look-up table correction, residual dispersion is measured and corrected in closed-loop as a background process while observations are underway, correcting PSF elongation in real-time. 

\section{Motivation and Background}
The Magellan Adaptive Optics eXtreme instrument (MagAO-X) is an Extreme-AO high-contrast imaging system for the 6.5-meter Magellan Clay Telescope located at Las Campanas Observatory in Chile.\cite{males2018magaox,males2022magaoxII} MagAO-X is optimized for work in the visible/NIR spectrum  (0.5-1 micron) and contains multiple broadband and narrowband imaging filters for its two EMCCD science cameras. For high-order wavefront sensing and control, MagAO-X uses a modulated Pyramid Wavefront Sensor and a 2040 actuator MEMS Deformable Mirror (DM) from Boston Micromachines Corporation. Science cases for MagaAO-X include the search for accreting protoplanets\cite{close2025wide,li2025discovery}, characterization of binary systems in optical/NIR wavelengths\cite{pearce2023hip,pearce2025five}, and imaging of resolved objects such as circumstellar disks\cite{kueny2026multiband}. MagAO-X is equipped with a number of coronagraphs for high-contrast observations and has an ultimate science goal of imaging nearby extrasolar planets in reflected light\cite{males2022magaoxII}. MagAO-X also acts as a pathfinder instrument for GMagAO-X, the first-light extreme AO system for the Giant Magellan Telescope (GMT). With a larger collecting area and higher resolution, GMagAO-X will target terrestrial, potentially habitable exoplanets.\cite{males2022conceptual}

\begin{figure} [t!]
\begin{center}
\begin{tabular}{c} %% tabular useful for creating an array of images 
\includegraphics[height=7cm]{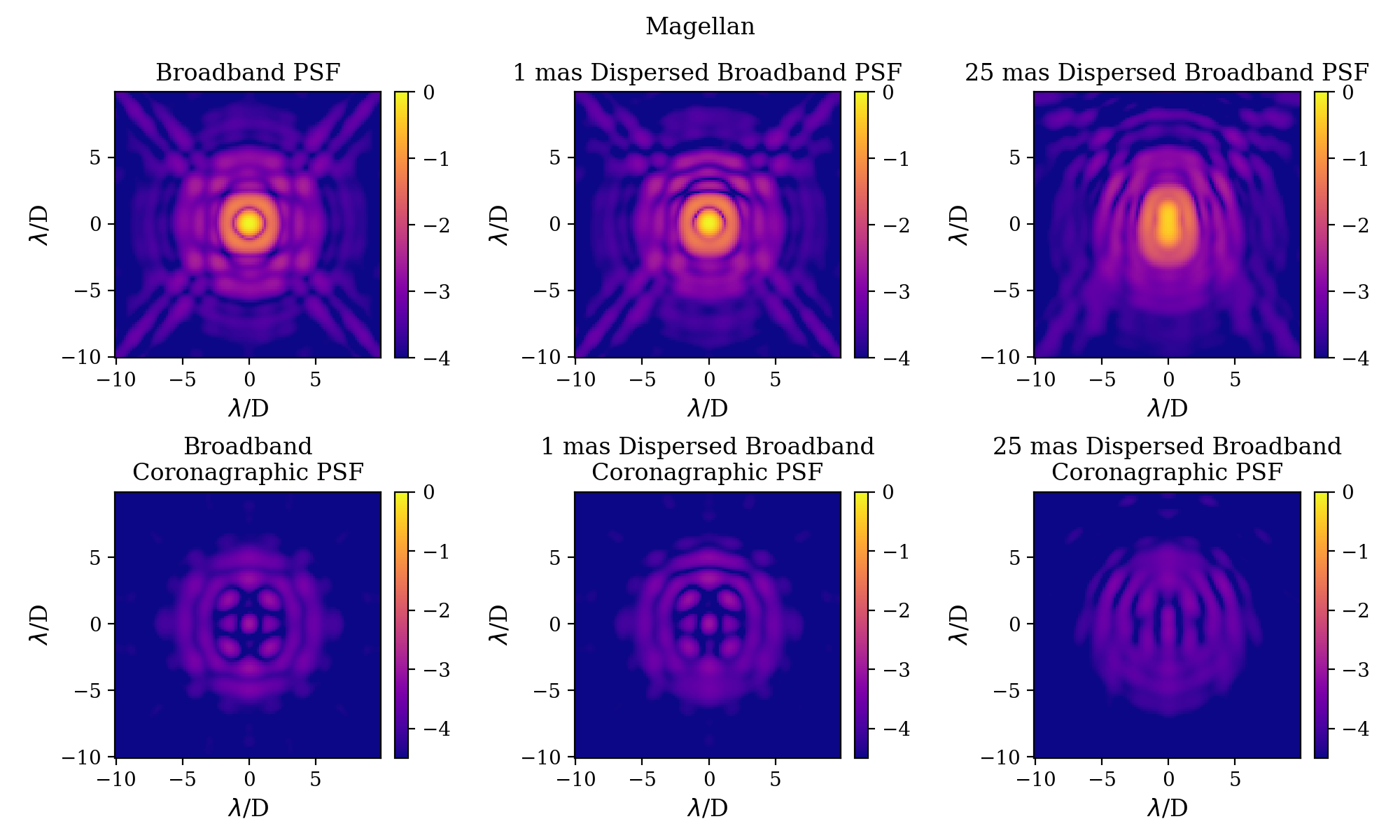} \\
\includegraphics[height=7cm]{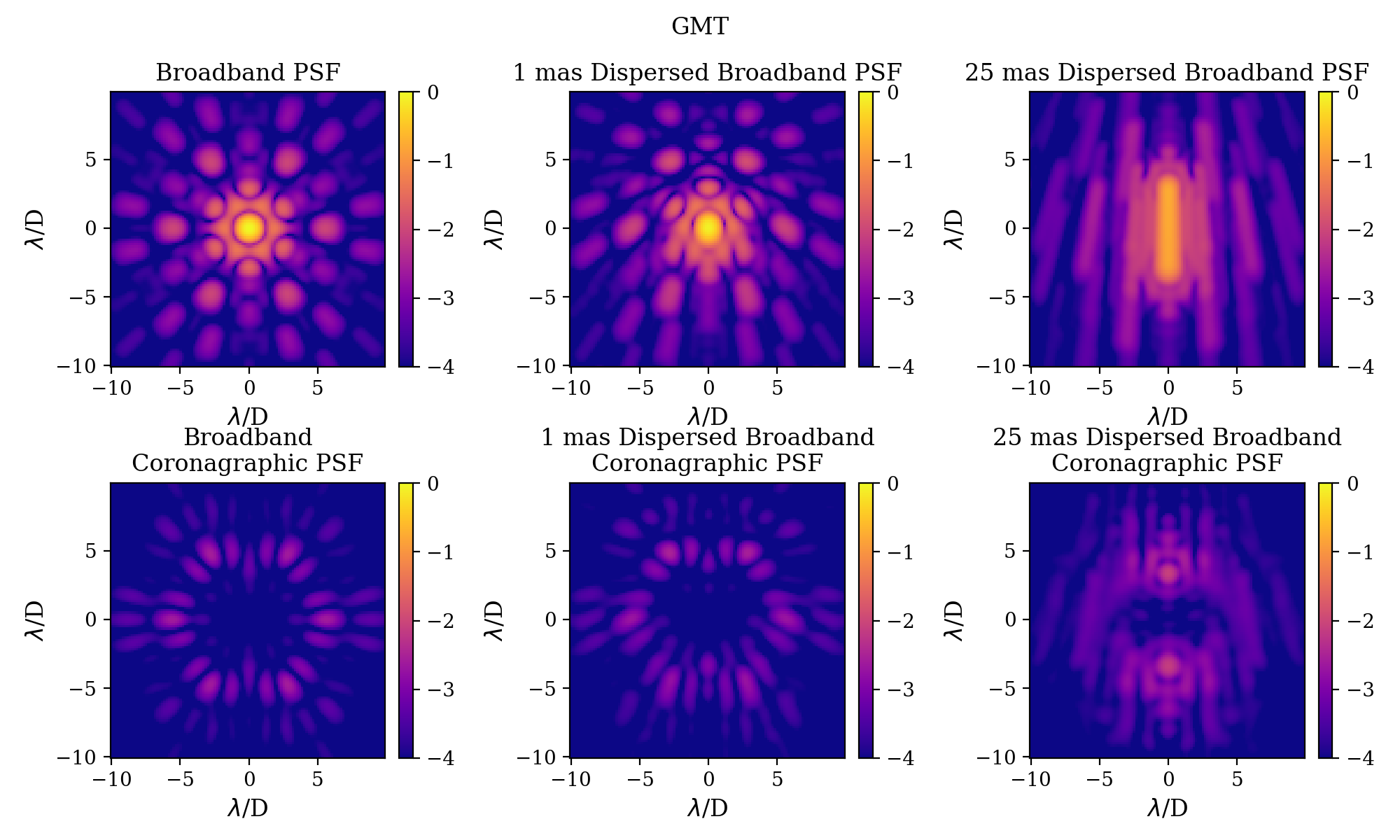}
\end{tabular}
\end{center}
\caption[coron] 
{ \label{fig:coron} 
Coronagraphic and non-coronagraphic broadband PSFs for the Magellan (top) and Giant Magellan (bottom) Telescopes simulated using hcipy\cite{por2018high}, with no residual wavefront error aside from the chromatic tip/tilt caused by dispersion. PSFs are simulated with wavelength 615 nm and bandwidth 110 nm, which matches the r' filter on MagAO-X. All colorbars are in logscale of the intensity, with each image normalized by the maximum intensity of its respective un-dispersed broadband PSF. A classical Lyot coronagraph simulation with a $3~\lambda/D$ radius focal plane mask was used to generate the coronagraphic PSFs. Because the GMT PSF is a smaller angular size, the same amount of dispersion in both systems affects the GMT PSF to a much greater extent. While 1 mas of dispersion on Magellan only slightly degrades the PSF quality, the GMT PSF is noticeably stretched, and the side lobes have become distorted. In the extreme case, with 25 mas of dispersion, the Magellan PSF is also noticeably elongated and distorted, and the GMT PSF leaks significantly around the top and bottom edges of the focal plane mask.
}
\end{figure}

For both instruments, deep contrast is critical to achieving their respective science goals. Coronagraph performance can be limited by a number of error terms, including residual wavefront error, quasi-static speckles, and the subject of this work: chromatic errors due to atmospheric dispersion\cite{guyon2005limits}. Figure \ref{fig:coron} shows the effect of uncorrected atmospheric dispersion, including elongation of the PSF, distortion of off-axis features, and coronagraph light leakage. Additionally, the increase in PSF size along the dispersion axis makes the coronagraph more sensitive to jitter along that same axis; vibrations in this direction are more likely to cause the central star PSF to fall off the focal plane mask and leak starlight into the final coronagraphic image.

GMagAO-X has a requirement that the residual dispersion be less than $<\frac{1}{20}\lambda/D$ in order to achieve its desired coronagraph performance. At a wavelength of 600 nm, this corresponds to $<0.25$ mas. The current standard of analytical models can be precise down to the 5 mas level for GMagAO-X's wavelength operating range \cite{wehbe2021sky}, which is not enough to meet this specification. To provide adequate dispersion correction to meet GMagAO-X requirements under varying atmospheric conditions, active dispersion control must be implemented\cite{Close2026GMagAOX}. Testing this on MagAO-X not only acts as a risk mitigation strategy for GMagAO-X but also improves the quality of MagAO-X scientific observations, which have been known to be impacted by residual dispersion.

\section{Methods}
\subsection{Satellite Spots}
This work uses a dispersion measurement method based on the work of Pathak et al.\cite{pathak2016high} that relies on the wavelength-dependence of diffraction to extract dispersion information from science images.  A two-dimensional sinusoidal diffraction grating placed in an intermediate pupil plane will generate four diffraction spots (plus lower-amplitude cross-terms) in the focal plane that are copies of the zero-order PSF. As with any grating, the spacing between the central PSF and the first-order diffracted satellite spots is proportional to the frequency of the grating and the wavelength of the incident light.

\begin{figure}[t]
    \centering
    \includegraphics[height=7cm]{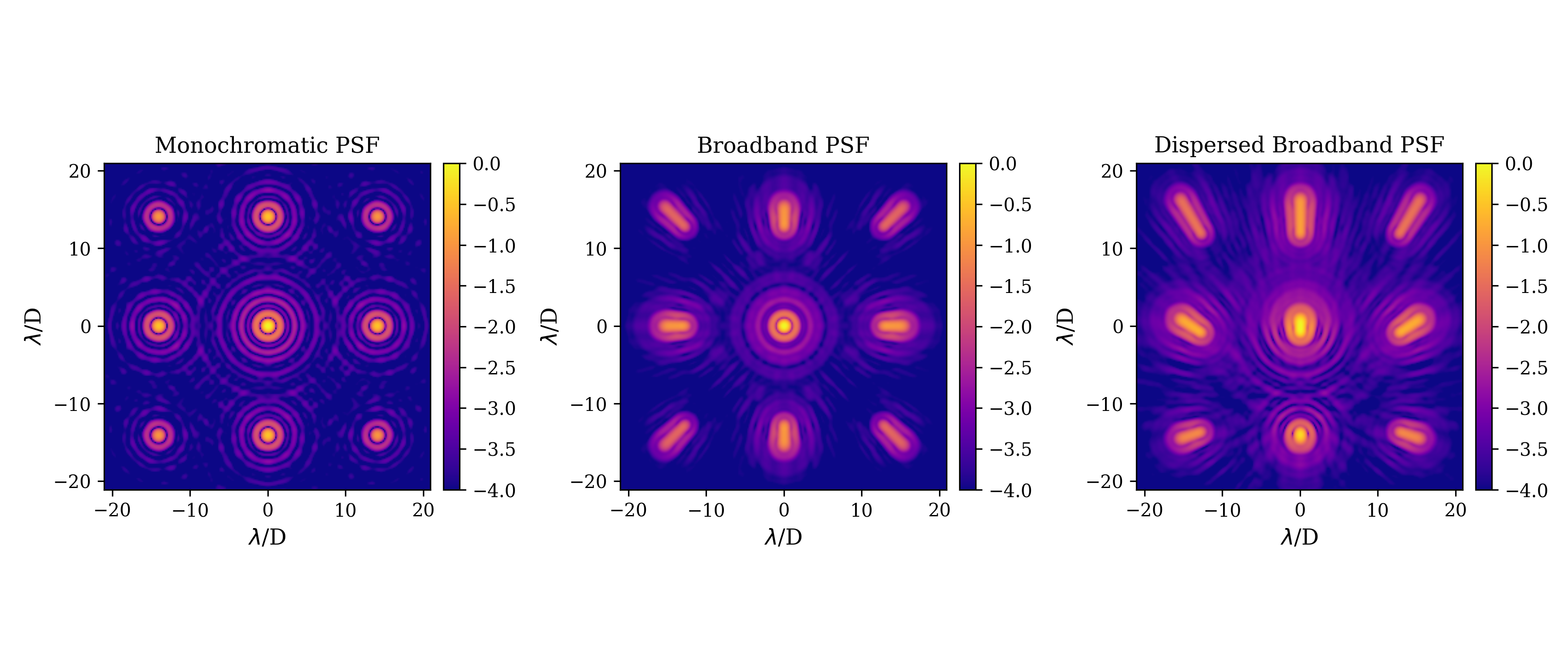}
    \caption{Simulated Magellan PSFs with satellite spots placed at 15 $\lambda/D$ using a sinusoidal diffraction grating. The four first-order satellite spots appear on the x- and y-axes, while the spots along the diagonals are second-order diffraction spots. Colorbars are in logscale of the intensity, normalized by the maximum intensity of the monochromatic PSF. The monochromatic satellite spots (left) are copies of the PSF, their intensity a function of the grating's first-order diffraction efficiency. The satellite spots in the nondispersed broadband image (center) are uniform and all point towards the zero-order PSF in the center. In the image with nonzero dispersion (right), some satellite spots are elongated while others are compressed, and they appear to point at some common center beneath the zero-order PSF. The zero-order PSF itself is also elongated in the direction of atmospheric dispersion, which is always along the y-axis for MagAO-X images.}
    \label{fig:dispersion_sparkles}
\end{figure}

Because broadband light can be considered as a superposition of monochromatic light of different wavelengths, broadband satellite spots appear as elongated versions of the PSF radiating from the zero-order PSF in the middle. However, when dispersion is introduced, the zero-order PSF becomes smeared along the direction of the dispersion. Each monochromatic satellite spot is now centered on a different zero-order PSF, and the broadband satellite spots appear to change shape and pointing direction, as shown in Figure \ref{fig:dispersion_sparkles}. For small amounts of dispersion, the change in satellite spot orientations is easier to observe and quantify than the elongation of the center PSF\cite{pathak2016high}. Residual aberrations from atmospheric turbulence or high-frequency vibrations could also cause the zero-order PSF to appear elongated but will not cause the satellite spots to rotate, making this a reliable way to measure dispersion without confounding it with other PSF imperfections. It is also important to note that on MagAO-X, the image derotator ensures that the pupil always remains in the same orientation\cite{hedglen2018optical}; as a result, the direction of atmospheric dispersion always remains the same in the focal plane. Residual dispersion caused by a mismatch between the ADC correction and the true atmospheric dispersion strength results in an elongation of the PSF that is always along the vertical direction in the focal plane. If the dispersion has a component along the orthogonal direction, it is due to other instrumental effects (See Section \ref{section:Instrumental Effects}).

Satellite spots can be generated actively by applying a 2D sinusoidal shape to the surface of the DM. The MagAO-X BMC 2k DM is capable of generating satellite spots at any separation out to $22~\lambda/D$. Modulating this sinusoid to produce incoherent satellite spots is regularly done during coronagraphic observations on MagAO-X to provide a photometric calibration for the absolute flux of the star\cite{mcewen2024sky}. Additionally, the print-through of the periodic actuator structure behind the face sheet of the DM creates satellite spots that are permanently in the focal plane of all MagAO-X observations. These passive satellite spots, separated from the zero-order PSF by about $47~\lambda/D$ in the focal plane, are at a much wider separation than the active calibrator satellite spots; consequently, they are more sensitive to small amounts of dispersion but are much fainter. Both active and passive satellite spots can be used to measure dispersion while scientific observations are ongoing without disrupting their quality. 

\subsection{Algorithm to Measure Dispersion}
\label{subsection:Algorithm}
 While this work uses satellite spots and their apparent pointing directions to measure the dispersion in an image, the algorithm that is used differs significantly from the previous approach at SCEx-AO. In an image with zero dispersion, all satellite spots will point at the center of the PSF. However, due to residual wavefront errors, and because we are often observing in coronagraphic mode, the exact center of the PSF is not always well known. Measuring the offset between where the satellite spots are pointed and the center of the PSF is therefore an unreliable way to determine the magnitude of dispersion in the image. Instead, we rely on the diffracted nature of the satellite spots to know that, when no dispersion is present in the image, pairs of opposing satellite spots will point at \textit{each other}. This negates the need for precise centroiding of the zero-order PSF. The offset in pointing between a single satellite spot and its pair on the opposite side of the zero-order PSF is used to measure the amount of residual dispersion left in the image.

 The dispersion measurement takes place in the science focal plane, after AO correction has taken place. Preliminary image processing is used to increase the signal of the satellite spots relative to the background. This includes taking an average of a stack of 10-20 images, masking the center, and low- and high-pass filtering to smooth noise and minimize the residual atmospheric speckle halo. The following procedure is then used to determine the amount of dispersion in the image:

\begin{enumerate}
    \item Isolate a satellite spot from the image using a small pixel window.
    \item Slice the new window by rows or by columns (depending on spot orientation) to obtain 1D cross-sections of the minor axis of the satellite spot. Determine which 1D slices contain the satellite spot by taking the ratio of the maximum to the minimum pixel value in a given slice and comparing it to a threshold value. Disregard slices that do not contain information from the satellite spot.
    \item Starting at the maximum pixel in each slice, search outward until the intensity experiences a local minimum. Set all values outside this "central core" to zero. Normalize the slice so that its peak is 1.
    \item Use a least-squares approach to fit a 1D Gaussian curve with mean $\mu_i$ and standard deviation $\sigma_i$ to the $i$th row/column. Store the $\mu_i$ value for each column.
    \item Plot the column indicies $i$ against their corresponding $\mu_i$ values. Calculate the slope of this line to determine the pointing direction of the satellite spot. Store this value.
    \item Repeat steps 1-5 for the corresponding satellite spot on the opposite side of the zero-order PSF.
    \item Find the difference in pointing direction between the two spots. If there is no dispersion, these numbers should be very close to zero. If there is dispersion, this ``pair offset" angle will be used as feedback to the controller to send the right amount of compensation.
\end{enumerate}

\begin{figure}
    \centering
    \includegraphics[height=6cm]{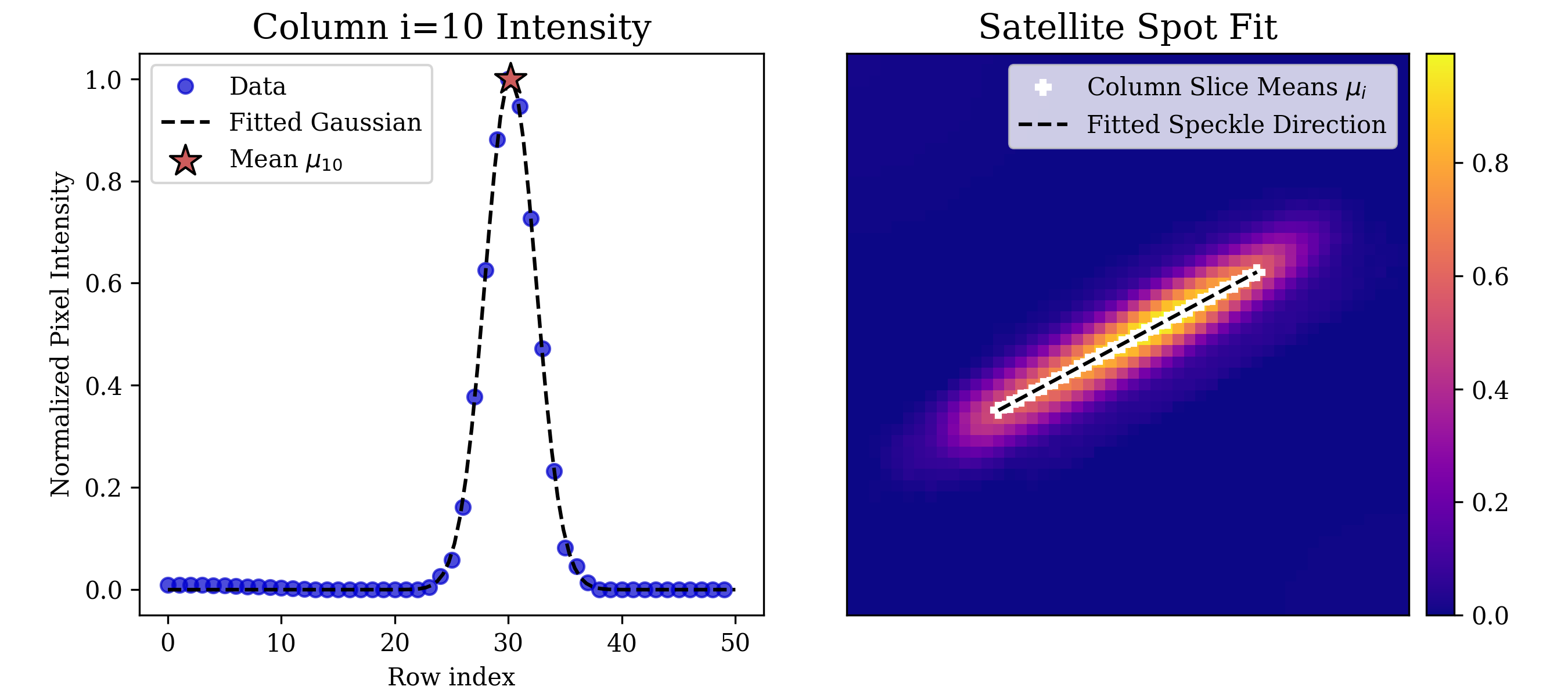}
    \caption{Illustration of the fitting process used to determine the pointing direction of a single satellite spot. The satellite spot (right) is a passively generated spot from a real image of an internal calibration source taken by one of MagAO-X's science cameras. The image is sliced into columns (left) and each column normalized before being individually fit to a Gaussian distribution, which helps to mitigate the effect of the spectrum on the fitting efficacy. The column means are fit with a linear slope, which determines the pointing direction of the satellite spot.}
    \label{fig:fitting}
\end{figure}

The fitting process is visualized in Figure \ref{fig:fitting}. Several factors influence the fidelity of the fitting, including the number of photons in the image, residual turbulence, the spectrum of the object, and the threshold value that determines whether a slice will be used or ignored. If there are multiple pairs of satellite spots at the same separation, the process can be repeated and the results averaged to obtain a more precise measurement. Due to the effect of satellite spot separation on their sensitivity to dispersion, satellite spots of different separations can contribute to the same measurement only after the offset angles have been converted into true dispersion units.

The output of the fitting algorithm is not a dispersion value in the typical $mas$ or $mas/\mu m$ units. In order to convert the pair offset angles to real units, a calibration was performed. The ADCs were counter-rotated by known amounts to produce residual dispersion in the focal plane, the elongation of the central PSF was measured alongside the satellite spot offset angles, and conversion factors were empirically calculated. This was repeated for each satellite spot separation of interest. Since the amount of residual dispersion left after the first-stage ADC lookup table correction is relatively small, linear conversion factors describe this behavior quite well. 

\subsection{Closed-loop Control}

\begin{figure}
    \centering
    \includegraphics[height=4cm]{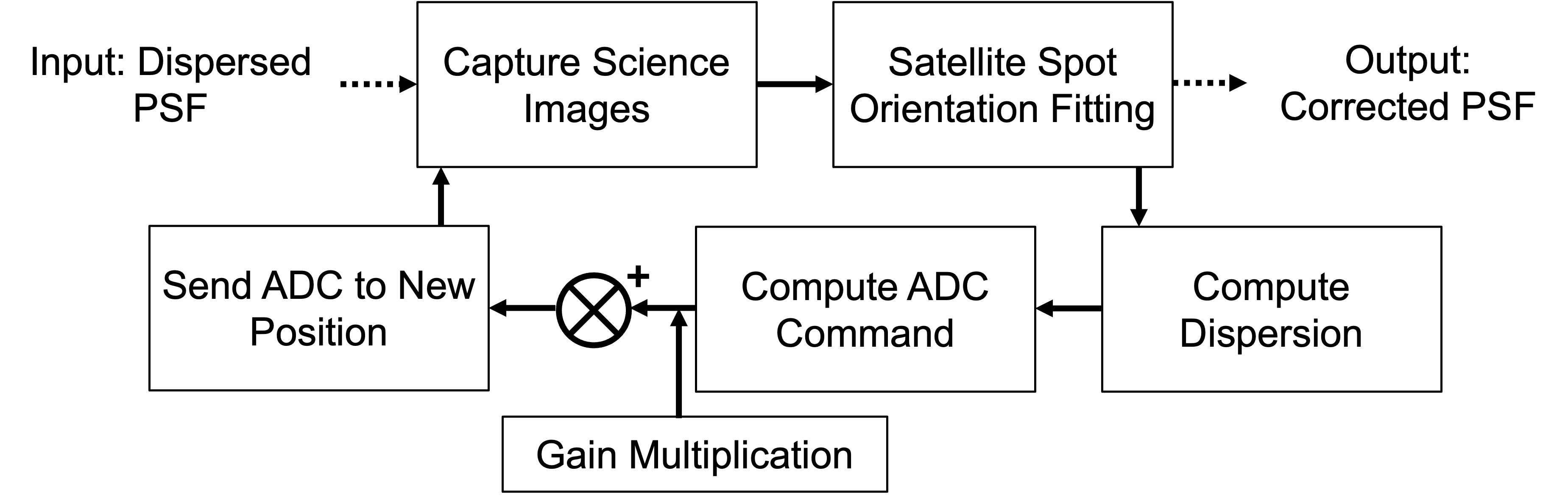}
    \vspace{0.7cm}
    \caption{Control loop diagram for active dispersion control. Images from the science cameras are taken and processed as described in Section \ref{subsection:Algorithm}. The satellite spots are used to extract the amount of dispersion in the image, and the ADC command is computed. A gain is applied, and the ADCs are sent to their new positions. The gain is a tunable parameter that can be lowered in poor atmospheric seeing conditions in order to promote control loop stability. 
    }
    \label{fig:control_loop}
\end{figure}

The control loop is depicted in Figure \ref{fig:control_loop}. Because the conversion factor between the satellite spot pointing offsets and the ADC angles has been well-calibrated, the procedure is relatively simple: the satellite spot fitting algorithm is run, a dispersion value is extracted, and a new ADC position is calculated. A scalar multiplicative gain is applied to improve control loop stability, and once the ADCs are sent to their new positions the process begins anew. A simple integrator controller is used and the loop converges on a solution that minimizes the dispersion in the image.

Since atmospheric dispersion evolves slowly compared to wavefront errors caused by turbulence,\cite{pathak2018sky} the frequency of the ADC control loop can be run much slower than MagAO-X's 2 kHz high-order wavefront sensing and control loop. This allows time to take multiple stacks of images and fit a dispersion value to each of them, exclude obvious outliers, and use the average measurement to calculate the new ADC position. Averaging over multiple measurements in this way increases robustness to fitting errors and greatly increases stability in the presence of strong turbulence, when the shape of the satellite spot is distorted by residual wavefront errors. Currently, the ADC control loop runs on MagAO-X at a speed of approximately 0.1 Hz.

\section{On-sky Demonstration}

\subsection{Nov. 2025 $\texorpdfstring{\tau}{Tau}$ Ceti}

\begin{figure}[b]
    \centering
    \includegraphics[height=6cm]{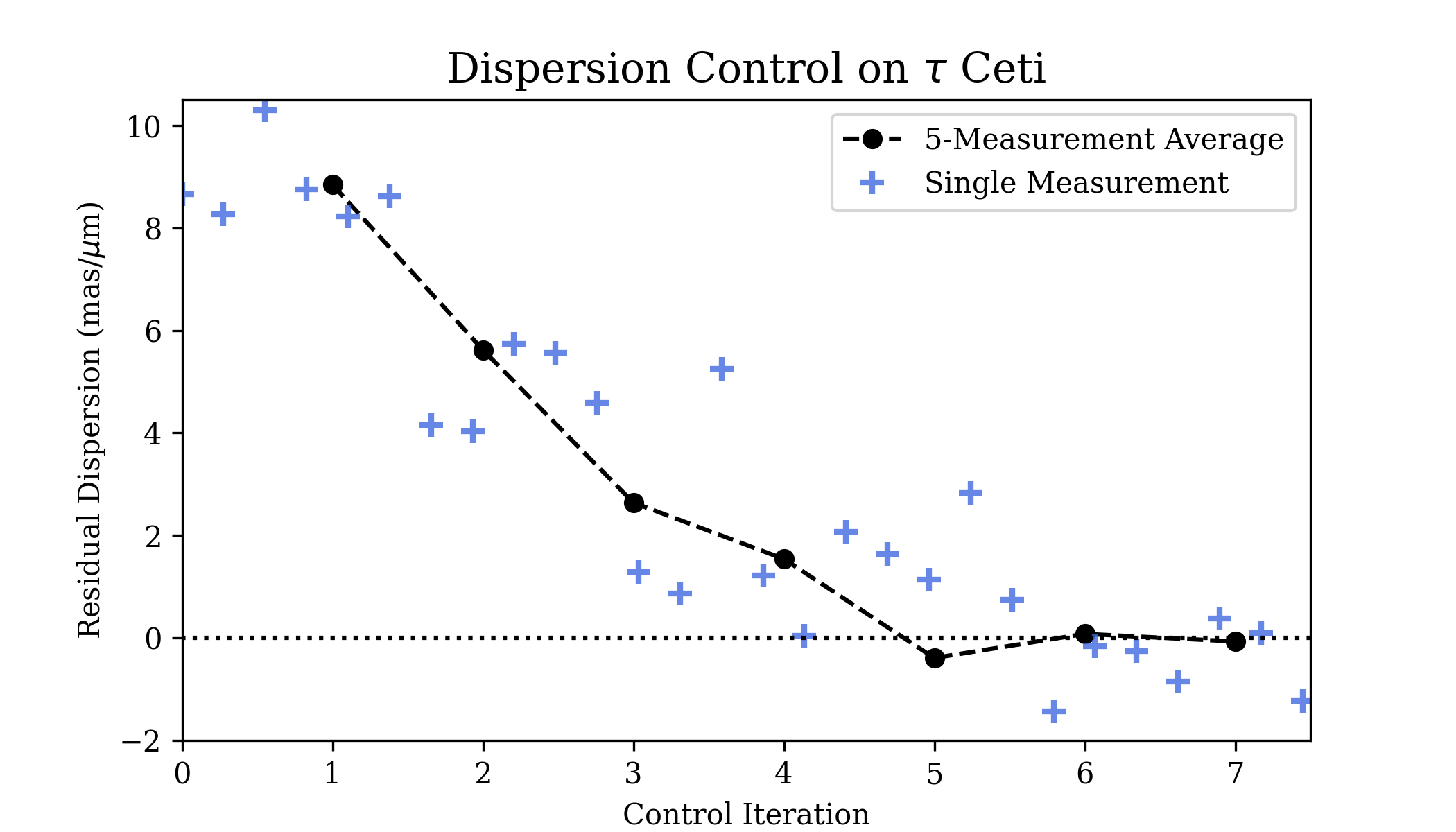}
    \vspace{0.5cm}
    \caption{Results from the Nov. 2025 $\tau$ Ceti ADC active control experiment. The amount of residual dispersion, as measured by the orientation of the satellite spots, is shown as a function of the number of iterations of active control. Dispersion is estimated in the focal plane five  times and the measurement is averaged before a command is sent to the ADC. The scalar gain value was 0.5. The algorithm takes only five iterations to settle to a steady-state with very little error, an average of 0.1 $mas/\mu m$ over the final three iterations.}
    \label{fig:tauceti}
\end{figure}

Closed-loop control of the MagAO-X ADCs using this algorithm was demonstrated for the first time on the night of 30 November 2025 on the object $\tau$ Ceti. This object was chosen because it has a magnitude of 2.41 in i-band, giving us bright satellite spots in the focal plane. Atmospheric seeing was variable, ranging between 0.62 and 1.2 arcseconds for the duration of the observations. Telescope elevation ranged from 51.9° - 53.1°, corresponding to an airmass range of 1.24 - 1.27. The observation was set up using MagAO-X's i-band filter with $\lambda=762$ nm and $\Delta\lambda=126$nm, and no coronagraph was used. Figure \ref{fig:tauceti} shows results from this experiment. The four outermost satellite spots, passively generated at 47 $\lambda/D$ by the 2k DM, were used for the dispersion measurement. In one test, a larger residual dispersion of 8.8 $mas/\mu m$ was induced by offsetting the ADCs by one degree from their nominal position as dictated by the lookup table. After five iterations of active control, the algorithm converges, and the residual dispersion magnitude was reduced to 0.1 $mas/\mu m$ along the vertical axis. 

\subsection{Mar. 2026 \texorpdfstring{$\alpha$}{alpha} Circini}

\begin{figure}
    \centering
    \includegraphics[height=6cm]{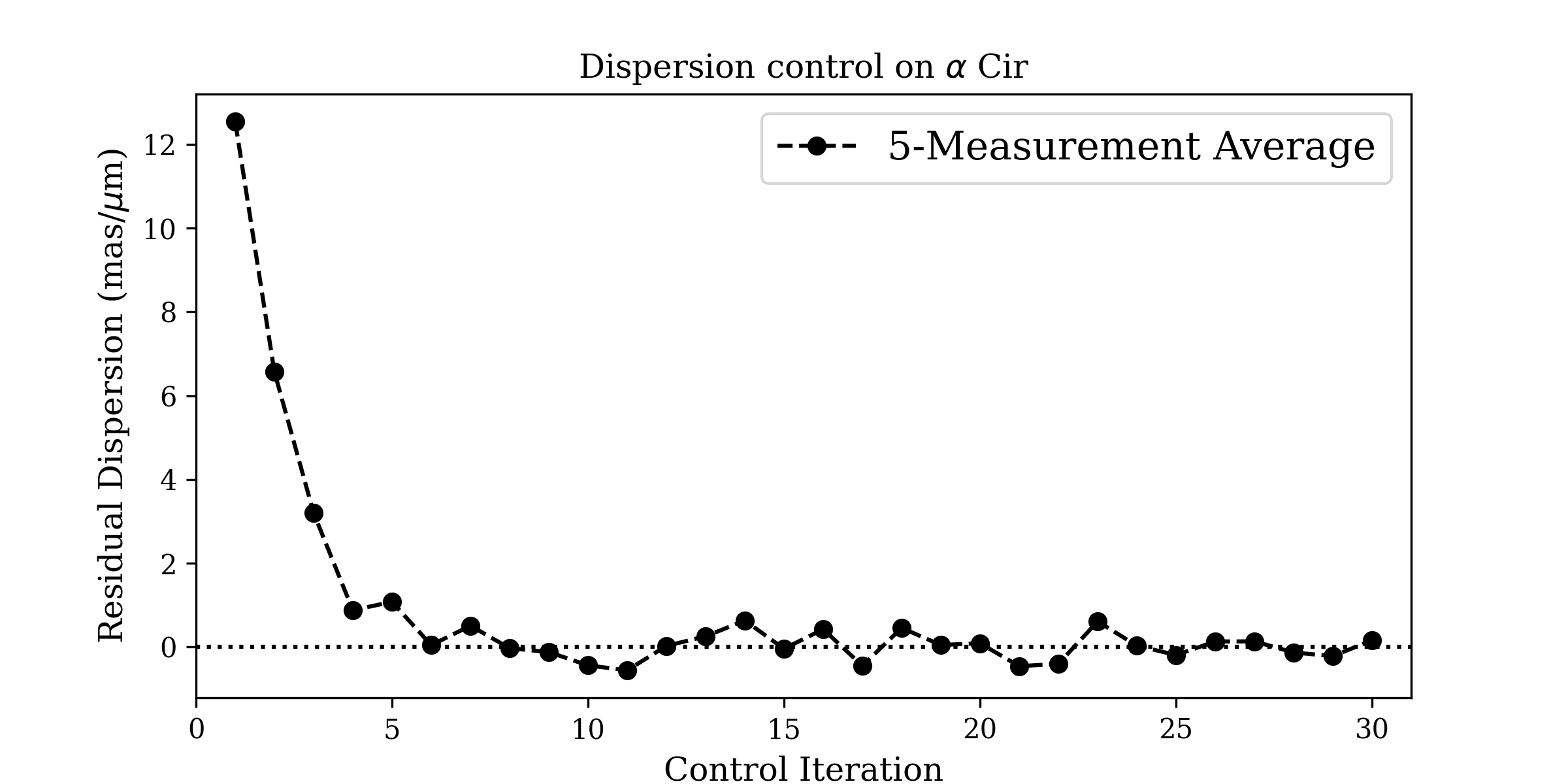}
    \vspace{0.5 cm}
    \caption{Results from the Mar. 2026 $\alpha$ Cir experiment to test the stability of the algorithm. Again, dispersion was estimated in the focal plane five times before a command was sent to the ADC, though the individual measurements are not shown this time to avoid crowding the plot. The control loop scalar gain was set to 0.5. After about six iterations, the controller settles into a steady-state. The magnitude of residual dispersion averages $0.26~mas/\mu m$ with a standard deviation of $0.20~mas/\mu m$ from iterations 7 - 30.
    }
    \label{fig:alphacir}
\end{figure}

Active ADC control was tried a second time on MagAO-X on the night of 31 March 2026. The object was $\alpha$ Circini, which has a magnitude of 2.86 in i-band, making it a similar brightness to $\tau$ Ceti. Atmospheric seeing was much better during these observations, ranging from 0.49 - 0.57 arcseconds. MagAO-X's large Lyot coronagraph was used, which has a focal plane mask radius of 4.1 $\lambda/D$ at 762 nm, the central wavelength of the i-band filter used for the observation. The telescope elevation angles were similar to the first set of experiments, varying from 52.4° to 52.6°, which corresponds to about 1.26 airmasses.

Similar to the previous experiment, a small offset in the ADCs was purposely introduced, and the loop was closed in order to correct it. For the test depicted in Figure \ref{fig:alphacir}, residual dispersion in the focal plane started at 12.5 $mas/\mu m$ and was reduced to 0.27 $mas/\mu m$ after six iterations. This time, the loop was run much longer to test the stability of the control algorithm. After convergence, the magnitude of residual dispersion along the vertical axis averaged $0.26~mas/\mu m$ with a standard deviation of $0.20~mas/\mu m$.

\section{Instrumental Dispersion Effects}
\label{section:Instrumental Effects}

\subsection{Axis Alignment}
\begin{figure}[ht]
    \centering
    \includegraphics[height=6cm]{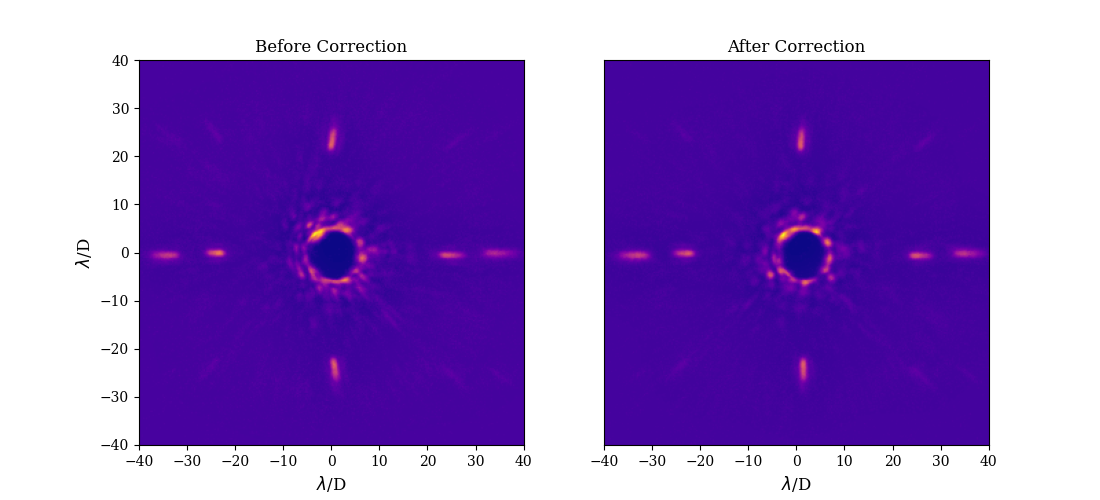} \\
    \includegraphics[height=6cm]{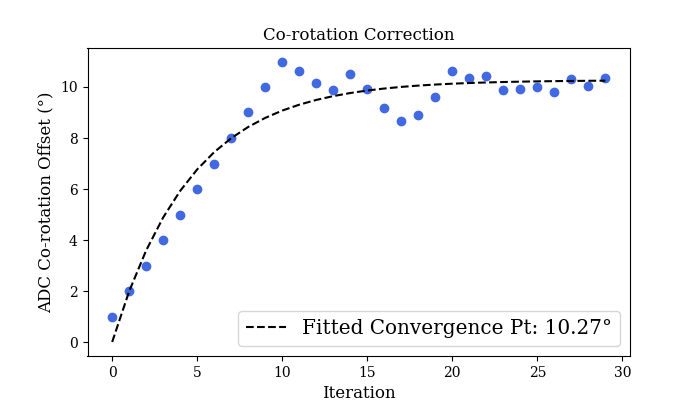}
    \vspace{0.5cm}
    \caption{Results from the March 2026 ADC experiments on $\alpha$ Circini. Top: two single-frame images of $\alpha$ Cir, before and after the ADC alignment correction was applied. Six satellite spots are seen: four active satellite spots at 15 $\lambda/D$ separation generated by modulating a sinusoid on the high-order DM, and two wider-separation satellite spots along the x-axis generated passively by the non-common path DM actuator pattern. The two satellite spots along the y-axis are clearly angled toward the left in the left image, while in the right image all four active satellite spots are of equal size and point towards the center of the coronagraphic PSF. Bottom: The active ADC control loop takes about 20 iterations to converge to a solution of 10.27°, the amount by which the ADC axis needs to rotate in order to line up with the atmospheric dispersion axis.}
    \label{fig:corotation}
\end{figure}

A recurring issue in MagAO-X's observations was a clear dispersion of the PSF along the horizontal axis, perpendicular to the expected direction of any residual atmospheric contribution. It was hypothesized that this was the result of a misalignment of the ADC two-prism group. Using the active ADC control algorithm, we co-rotated the ADCs as a pair (rather than counter-rotating them, like in the previous two on-sky demonstrations) to eliminate this horizontal component. Figure \ref{fig:corotation} shows the results of this experiment, performed during the same observation block in March 2026 on $\alpha$ Cir, demonstrating that the MagAO-X ADC axis was misaligned from the true atmospheric dispersion axis by 10.27°. With this correction applied, the horizontal component of the dispersion is reduced significantly.

\subsection{Lateral Color}

\begin{figure}
    \centering
    \includegraphics[height=5.5cm]{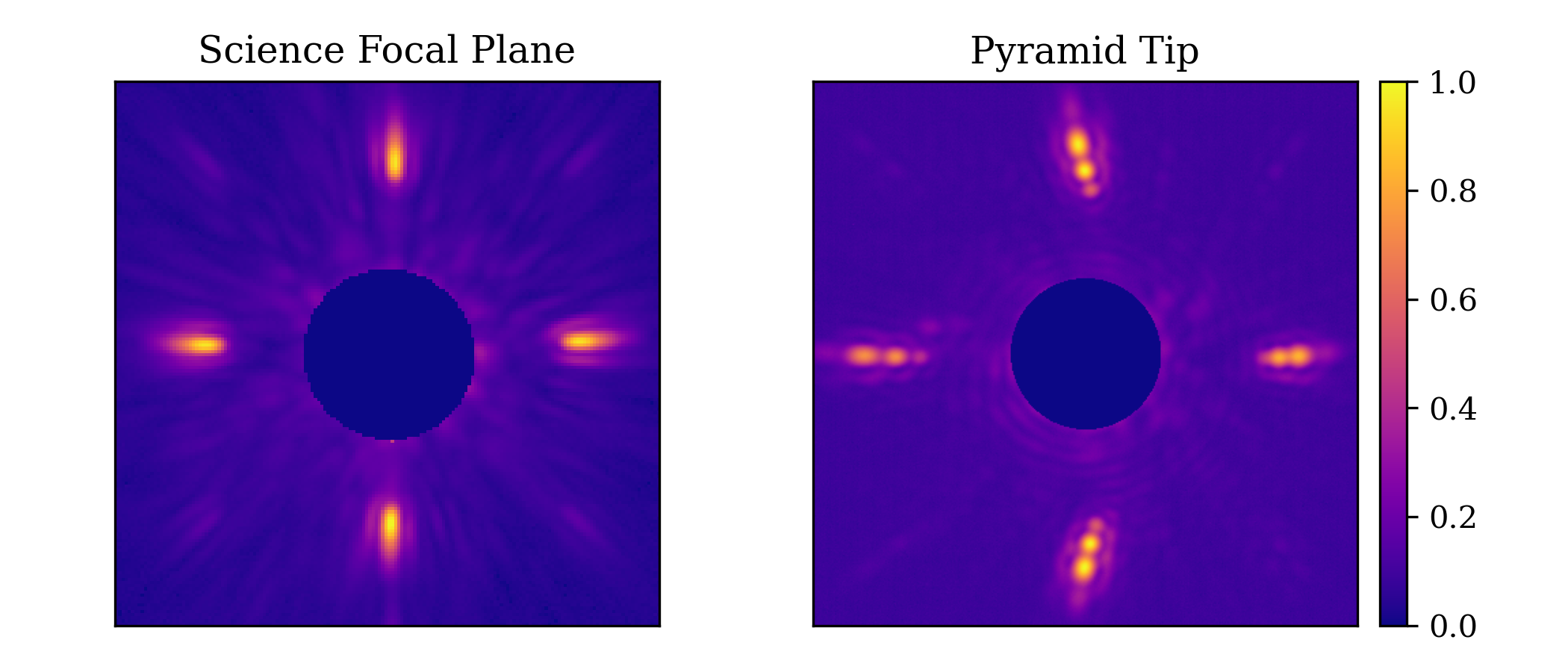}
    \caption{Non-common path dispersion between the wavefront sensing and the science paths. These two images were taken simultaneously, the left in the final science focal plane and the right in the plane of the tip of the pyramid. Intensities have been normalized to the peak of each image, and the zero-order PSF has been masked. In the left image, the spectrum of the pellicle beamsplitter that sends the light into the pyramid tip camera causes the spectrum of those satellite spots to have absorption lines. The satellite spots are all pointed towards the center in the science image, but the top and bottom satellite spots are clearly deviated to the right at the pyramid tip, indicating the presence of dispersion.}
    \label{fig:ncp dispersion}
\end{figure}

After the dispersion axis was corrected, another issue was discovered upon examining simultaneously the PSF in the science focal plane and the PSF in the focal plane at the tip of the pyramid wavefront sensor. Figure \ref{fig:ncp dispersion} shows that there is a clear non-common path dispersion between the two. This effect can only be caused by refractive optics internal to the system, and results in chromatic shearing of the pyramid pupils, a term that contributes to wavefront reconstruction error\cite{correia2020hcipwfs}. Unfortunately, the amount of pupil shearing in the wavefront sensing path is a direct tradeoff with the amount of dispersion in the science path, since the ADCs are upstream of the beamsplitter that splits the two paths. This tradeoff is especially noticeable because MagAO-X has an operational mode that uses an un-modulated pyramid sensor and a neural network reconstructor\cite{landman2024making,landman2025making}. When the ADCs correct dispersion in the science plane, they increase dispersion at the pyramid, creating an effect the neural network has not been trained for and which it cannot correct using the DM. To eliminate this tradeoff, a compensating refractive optic will be placed in the wavefront sensing path in the future.

\section{Conclusions and Further Work}
Active ADC control has been shown to reduce residual dispersion in MagAO-X on-sky science images. This is the first time that active ADC control has been demonstrated in a single waveband and in the visible-NIR spectrum, and it works with either actively or passively generated satellite spots. Active ADC control on MagAO-X lays the foundation for high-precision dispersion control on GMagAO-X, which will be necessary to reach coronagraph performance specifications. Using a 6.5-meter telescope, we were able to achieve sub-$mas/\mu m$ residual dispersion levels and approach the performance requirement of GMagAO-X. However, these demonstrations were performed only on bright targets, and more simulated and on-sky experimentation will be necessary to determine the magnitude limit of this method.

\acknowledgments % equivalent to \section*{ACKNOWLEDGMENTS}       
 
MagAO-X was developed with support from the NSF MRI Award No. 1625441. The Phase II upgrade program is made possible by the generous support of the Heising-Simons Foundation. MagAO-X uses the CACAO software package, which is supported by NSF Award No. 2410616. The GMagAO-X project is grateful for support from the University of Arizona Space Institute for the preliminary design phase and to the GMT for supporting final design.

% References
\bibliography{report} % bibliography data in report.bib
\bibliographystyle{spiebib} % makes bibtex use spiebib.bst

\end{document}